\documentclass[allclo]{FBSart}
\usepackage{amsfonts}
\usepackage{amssymb}
\usepackage{epsfig}
\usepackage{graphicx}
\title{Lepton Scattering off Few-Nucleon Systems at Medium and High Energies}
\author{\underline{C. Ciofi degli Atti}\instnr{1}\thanks{\textit{E-mail address:}\,ciofi@pg.infn.it},
L. P. Kaptari\instnr{2}, H. Morita\instnr{3}}
\instlist{Department of Physics, University of Perugia and Istituto
       Nazionale di Fisica Nucleare, Sezione di Perugia, Via A. Pascoli, I-06123, Italy\and
       Bogoliubov Lab. Theor. Phys., 141980,
         JINR, Dubna, Russia\and
       Faculty of Social Information, Sapporo Gakuin University,
        11 Bunkyo-dai, Ebetsu-shi, Hokkaido 069-8555, Japan}
\runningauthor{}
\runningtitle{}
\usepackage[figuresright]{rotating}

\newcommand{\AmS}{{\protect\the\textfont2  A\kern-.1667em\lower.5ex\hbox{M}\kern-.125emS}}

\def\bq{{\mbox{\boldmath$q$}}}

\def\bp{{\mbox{\boldmath$p$}}}

\def\br{{\mbox{\boldmath$r$}}}
\def\b0{{\mbox{\boldmath$0$}}}

\def\bq{{\mbox{\boldmath$q$}}}

\def\bp{{\mbox{\boldmath$p$}}}
     \font\tenbifull=cmmib10 scaled 1200 
     \font\tenbimed=cmmib9
     \font\tenbismall=cmmib7
\mathchardef\bbkappa="7114
\mathchardef\bbrho="711A
\mathchardef\bbsigma="711B
\mathchardef\bbtau="711C
\mathchardef\bbvarrho="7125
\mathchardef\bbvarsigma="7126
\mathchardef\bbxi="7118

\def\br{{\mbox{\boldmath$r$}}}
\def\b0{{\mbox{\boldmath$0$}}}

\def \b #1{ {\bf #1}}
\newcommand{\be}{\begin{eqnarray}}
\newcommand{\ee}{\end{eqnarray}}

\def \b #1{ {\bf #1}}

\def \b #1{ {\bf #1}}
\begin{document}
\maketitle
\begin{abstract}
The interpretation of recent Jlab experimental data
on the  exclusive process A(e,e'p)B off few-nucleon systems
 are  analyzed  in terms of
realistic nuclear wave functions  and Glauber multiple scattering
theory, both in its original form and within a generalized eikonal
approximation. The relevance of the exclusive process $^4He(e,e'p)^3H$  for possible
 investigations of QCD effects is illustrated.
\end{abstract}

Exclusive and  semi-inclusive lepton scattering off nuclei in the
quasi elastic region,  plays a relevant role in nowadays hadronic
physics  for the following main reasons: i) due to the wide
kinematical range available by present experimental facilities,
non trivial information on nuclei (e.g. nucleon-nucleon (NN)
correlations) can be obtained; ii) the mechanism of propagation of
hadronic states in the medium can be investigated in great
details; iii) at high energies QCD related effects (e.g. color
transparency effects) might be investigated. At medium and high
energies the propagation of a struck hadron in the medium  is
usually treated within the Glauber multiple scattering approach
(GA) \cite{glauber},  which has been applied with great success to  hadron
scattering off nuclear targets. However, when the hadron is
created inside the nucleus, as in a process $A({\it l},{\it
l}'p)X$, various improvements of the original GA have been
advocated. Most of them are based upon a Feynman diagram
reformulation of GA; such an  approach, developed long ago for the
treatment of hadron-nucleus scattering \cite{gribov}, has been
recently generalized to the process $A({\it l},{\it l}'p)X$
\cite{Misak05}, demonstrating that in particular kinematical
regions appreciable  deviations from GA are expected. The merit of
the  approach, based upon a generalized eikonal
 approximation  (GEA), is that  the   {\it frozen approximation}, common to
GA, is partly removed by taking
 into account the excitation energy of the  $A-1$ system;
  this results in a correction term to the standard
 profile function of GA,  leading   to an additional contribution to the
  longitudinal component of the missing momentum.
The GEA has recently  been applied \cite{CiofiRev,nofac} to a systematic calculation of the
 two-body (2bbu)  and three-body  (3bbu)  break up channels of
 $^3He$ electro-disintegration
 using realistic three-body  wave functions
\cite{Kiev} and  two-nucleon interactions (AV18) \cite{AV18}; the
two-body break up channel  $^3He(e,e'p)^2H$ has also been
considered within GEA  in Ref. \cite{rocco}, obtaining
results consistent with Ref. \cite{CiofiRev}. $A-1$.

 In GEA the final state wave function has  the following form
\be \Psi_f^*(\br_1,\,.\,.\,.\br_A)={\hat{\cal A}}
S_{GEA}(\br_1,\,.\,.\,.\br_A)e^{-i \bp\,\br_1} e^{-i
{\bf P}_{A-1} {\bf R}_{A-1}} \Phi_{A-1}^*(\br_2,\,.\,.\,.\br_A)
\label{states}
\ee
\noindent where
 $ {\mathcal S}_{GEA}=\sum \limits_{n=2}^{A}
{\mathcal S}_{GEA}^{(n)}$  generates the  final state interaction (FSI) between the struck particle and the
$A-1$ nucleon system;  in Eq. (\ref{states})
$n$ denotes  the order of multiple scattering, with the  single
scattering term ($n$=1) given by
\begin{eqnarray}
{\mathcal
S}_{GEA}^{(1)}(\br_1,\,.\,.\,.\br_A)=1-\sum\limits_{i=2}^A
\theta(z_i-z_1){\rm e}^{i\Delta_z (z_i-z_1)} \Gamma (\b
{b}_1-\b{b}_i) \label{essesingle}
\end{eqnarray}
\noindent where $\Gamma({\bf b})
={\sigma_{NN}^{tot}(1-i\alpha_{NN})} \cdot exp\,[-{\bf
b}^2/2b_0^2]/[4\pi b_0^2]$ is the usual Glauber profile function
and $\Delta_z = ({q_0}/{|{\b q}|}) E_m$, $E_m$ being the removal
energy related to the excitation energy of $A-1$; due to the presence of
$\Delta_z$  the frozen approximation is partly removed; note that
when $\Delta_z=0$, the usual GA is recovered (the expression of
the $n$-th order contribution ${\mathcal S}_{GEA}^{(n)}$ is given
in Ref. \cite{Misak05}).

 Within the factorization approximation
(FA), the diagrammatic approach leads to the following expression
for cross section
\begin{equation}
\frac{d^6 \sigma}
  {d \Omega ' d {E'} ~ d^3{\b p}_m} = {\mathcal
K}\sigma_{ep}P_A^{FSI}(\mathbf{p}_m,E_m),
\label{eq:crosssection}
\end{equation}
where ${\mathcal K}$ is a kinematical factor, $\sigma_{ep}$ the
electron-nucleon cross section,
 ${\bf p}_m$ = ${\b q} - {{\b p}}$ and  $E_m$ the missing momentum and energy, respectively,
  ${\b p}$  the momentum of the detected nucleon and, eventually, $P_A^{FSI}$ the distorted
  spectral function.

\begin{figure}[htb]
\centerline{\epsfysize=8.0cm\epsfbox{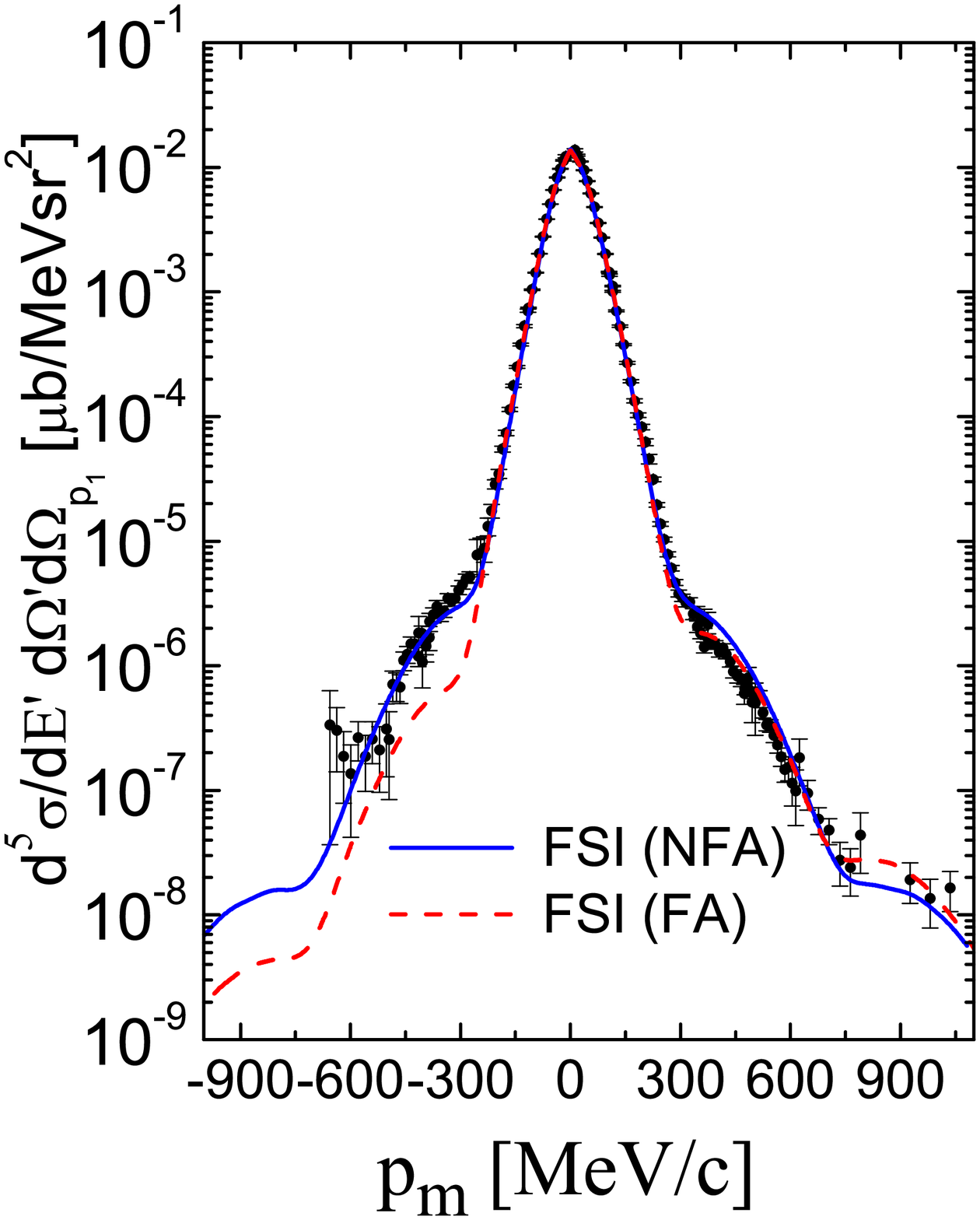}
    \epsfysize=6.5cm\epsfbox{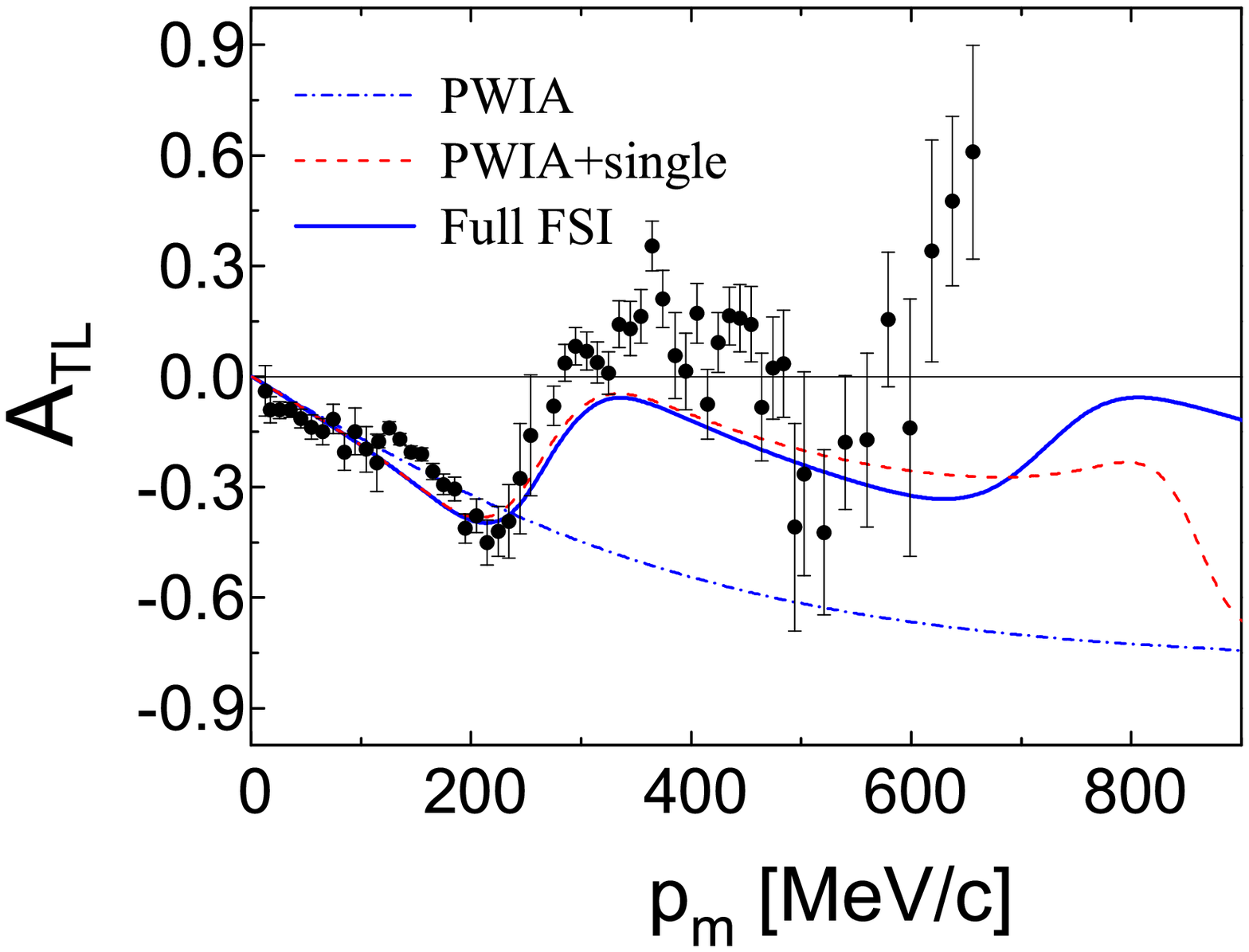}}
\caption{LEFT: the differential cross section of the process
$^3He(e,e^\prime p)^2H$ calculated taking into account FSI within
the non factorized (FSI(NFA)) and factorized (FSI(FA)) approaches.
Experimental data from Ref. \cite{E89044}. (After Ref.
\cite{nofac}).\,\, RIGHT: the transverse-longitudinal asymmetry
for the process $^3He(e,e^\prime p)^2H$. {\it Dot-dash}: PWIA;
{\it dash}: PWIA plus single rescattering FSI; {\it full}: PWIA
plus single and double rescattering FSI
 (three-body wave function from \protect\cite{Kiev},  $AV18$ interaction
     \protect\cite{AV18}). Experimental data from Ref. \cite{E89044}. (After Ref. \cite{nofac})}
 \label{Fig1}
\end{figure}
 If the FA is relaxed,  the  differential cross section assumes  the
 following form
 \be
 \frac{d^6 \sigma}
  {d \Omega ' d {E'} ~ d^3{\b p}_m} = \sigma_{Mott} ~ \sum_i ~
V_i ~ W_{i}^A( \nu , Q^2, {\b p}_m, E_m)
 \label{cross}
 \ee
 where $i \equiv\{L,
T, TL, TT\}$, and $V_L$, $V_T$, $V_{TL}$, and $V_{TT}$ are
well-known kinematical factors. A non factorized approach  (NFA) thus
requires  therefore the evaluation of the various response
functions $W_i$'s.  The cross sections of the processes
$^2He(e,e'p)n$, $^3He(e,e'p)^2H$, $^3He(e,e'p)(np)$, and
$^4He(e,e'p)^3H$  have been calculated in \cite{CiofiRev,nofac,
CKM} within a  parameter-free approach based upon realistic
  two-, three-, and four-body wave functions. In Fig. \ref{Fig1} the factorized and non
factorized cross sections of the 2bbu process
  $^3He(e,e^\prime p)^2H$ are compared with recent Jlab experimental data
  \cite{E89044}. The results presented in Fig. \ref{Fig1} clearly show that treating FSI within
the FA   is a poor approximation for "negative"  ( left, $\phi = 0$)
values of the missing momentum,  unlike what happens for  "positive"
(right, $\phi =\pi$) values (here $\phi$  is  the azimuthal  angle of the detected
 proton, with respect to the
 momentum transfers $\bq$). In spite of the good agreement provided by the NFA,
quantitative disagreements with experimental data still persist at
$\phi = 0$,  in particular  in the region around
 $|{\b p}_m|\simeq 0.6-0.65\,\, GeV/c$.
The origin of such a disagreement can better visualized by
analyzing the left-right asymmetry
\be
A_{TL}
=\frac{d\sigma(\phi=0^o)-d\sigma(\phi=180^o)}{d\sigma(\phi=0^o)+d\sigma(\phi=180^o)}.
\label{atl}
\ee
  It is well known that when  the explicit expressions of $V_i$ and $W_i^A$ are used in Eq.
  (\ref{atl})
 the
   numerator  is proportional to the transverse-longitudinal response $W_{TL}$, whereas the
 denominator does not contain $W_{TL}$ at all, which means that
 $A_{TL}$ is a measure of   the relevance of the transverse-longitudinal
 response relative to the other responses. The experimental \cite{E89044} and theoretical
 \cite{nofac}
  asymmetries are  presented in Fig. \ref{Fig1}, which clearly
  shows that at high values of the
   missing momentum
    the theoretical calculation cannot explain the experimental data. The reason for such a failure, which
    is common to  many approaches,  is at present under investigation.

\begin{figure}[!htp]
\centerline{\epsfysize=8.0cm\epsfbox{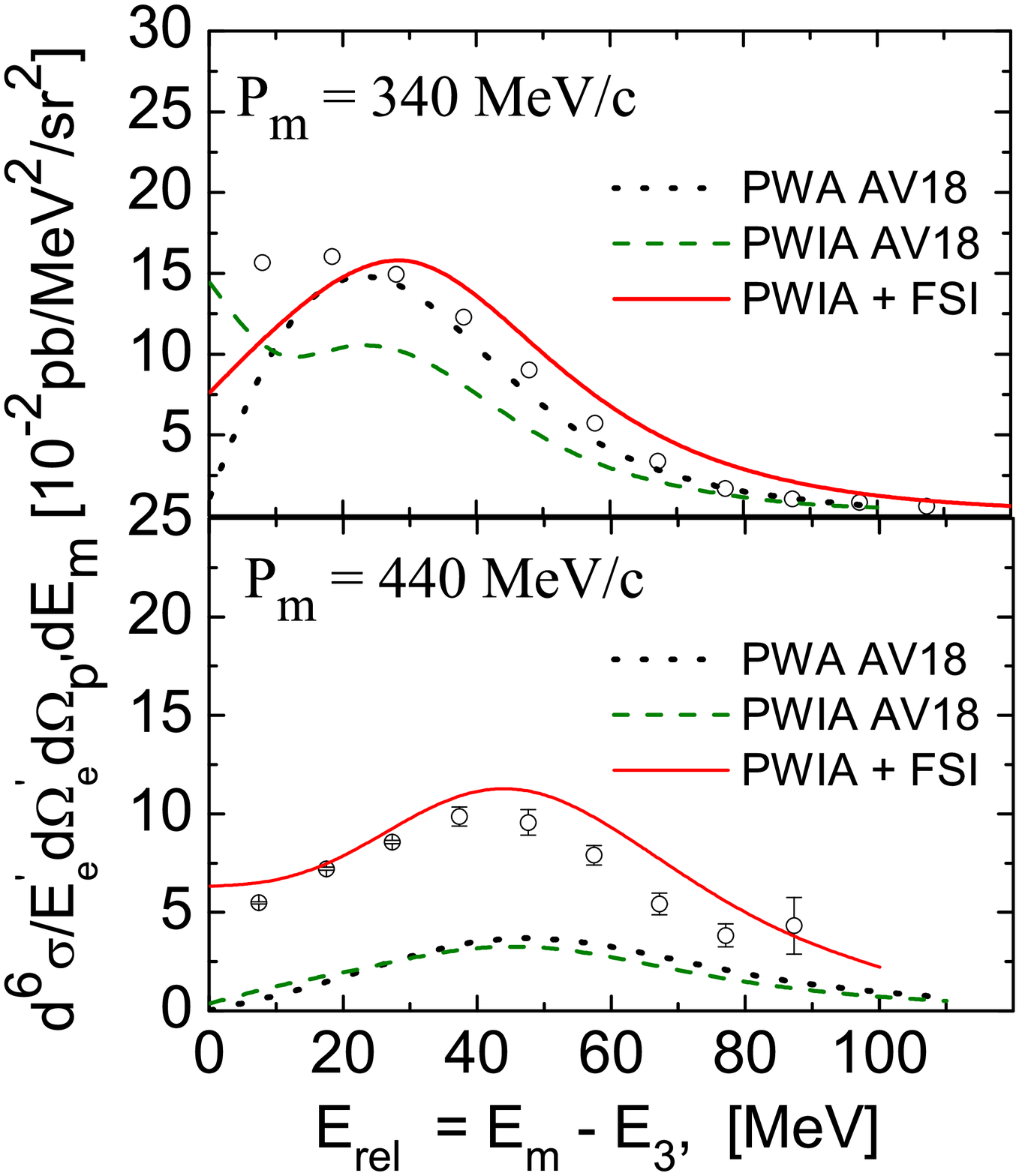}
    \hspace{-1cm}\epsfysize=8.0cm\epsfbox{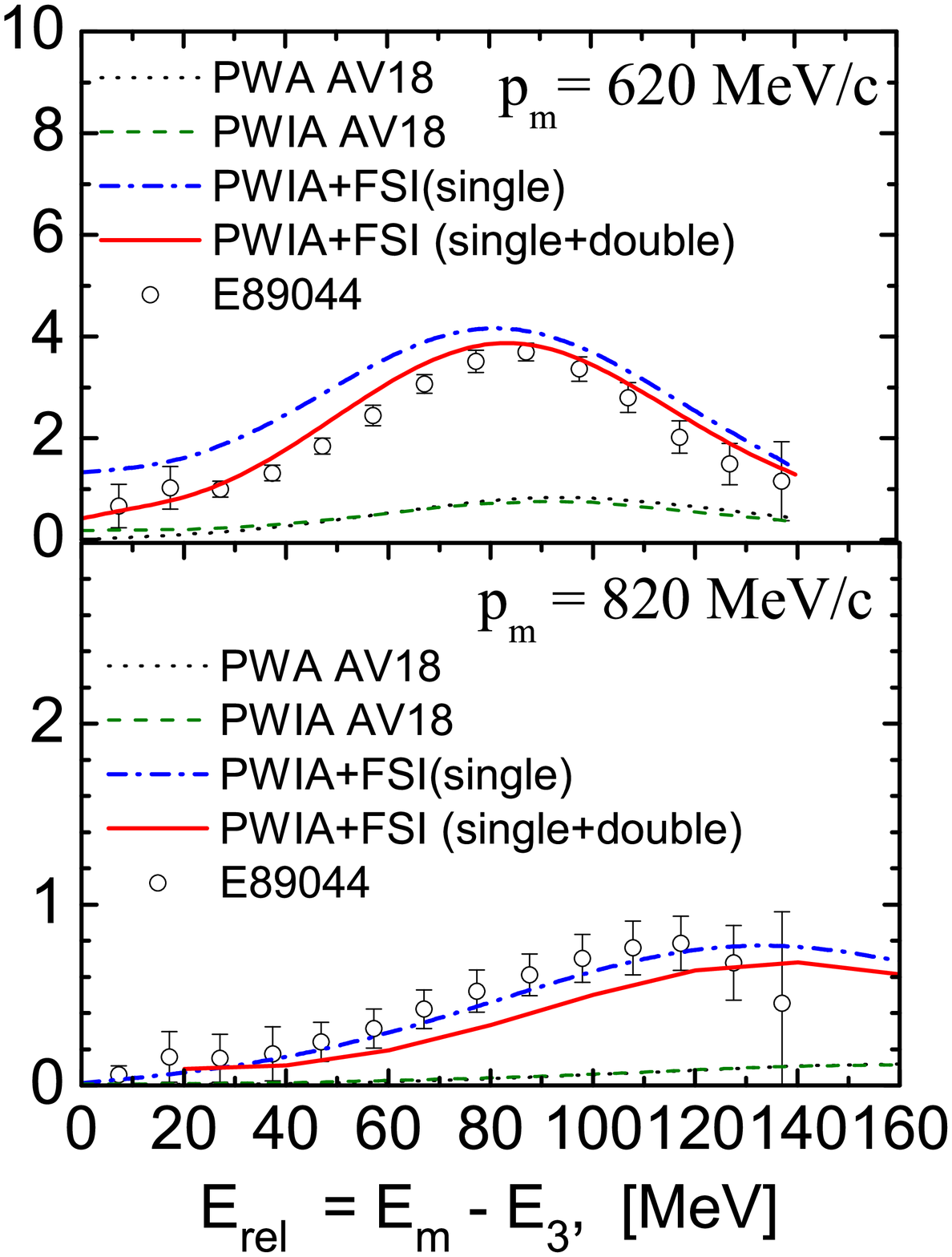}}
\caption{The differential cross section   of  the 3bbu channel $^3He(e,e'p)(np)$
  plotted, for fixed values of $p_m$, {\it vs}
the excitation energy of the two-nucleon system in the continuum
  $E_{rel}={\b {t}^2}/{M_N}$ = $E_2^f = E_m- E_{3}$.  The
curves labeled $PWA$ do not include any FSI;
   the dashed curves correspond to the
PWIA ; the dot-dashed curve include the FSI with single
rescattering;
  the full curves
include both single and double rescattering
     (three-body wave function from \protect\cite{Kiev},  $AV18$ interaction
     \protect\cite{AV18}). Experimental data from Ref. \cite{E89044}. (After Ref. \cite{CiofiRev})
     }
  \label{Fig2}
\end{figure}

Concerning the  3bbu channel calculation,  theoretical results are
presented in Fig. \ref{Fig2};  an overall good agreement with the
experimental data can be achieved,
 provided the large effects of the final state interaction are  taken into account.

\begin{figure}[htb]
\centerline{\epsfysize=8.0cm\epsfbox{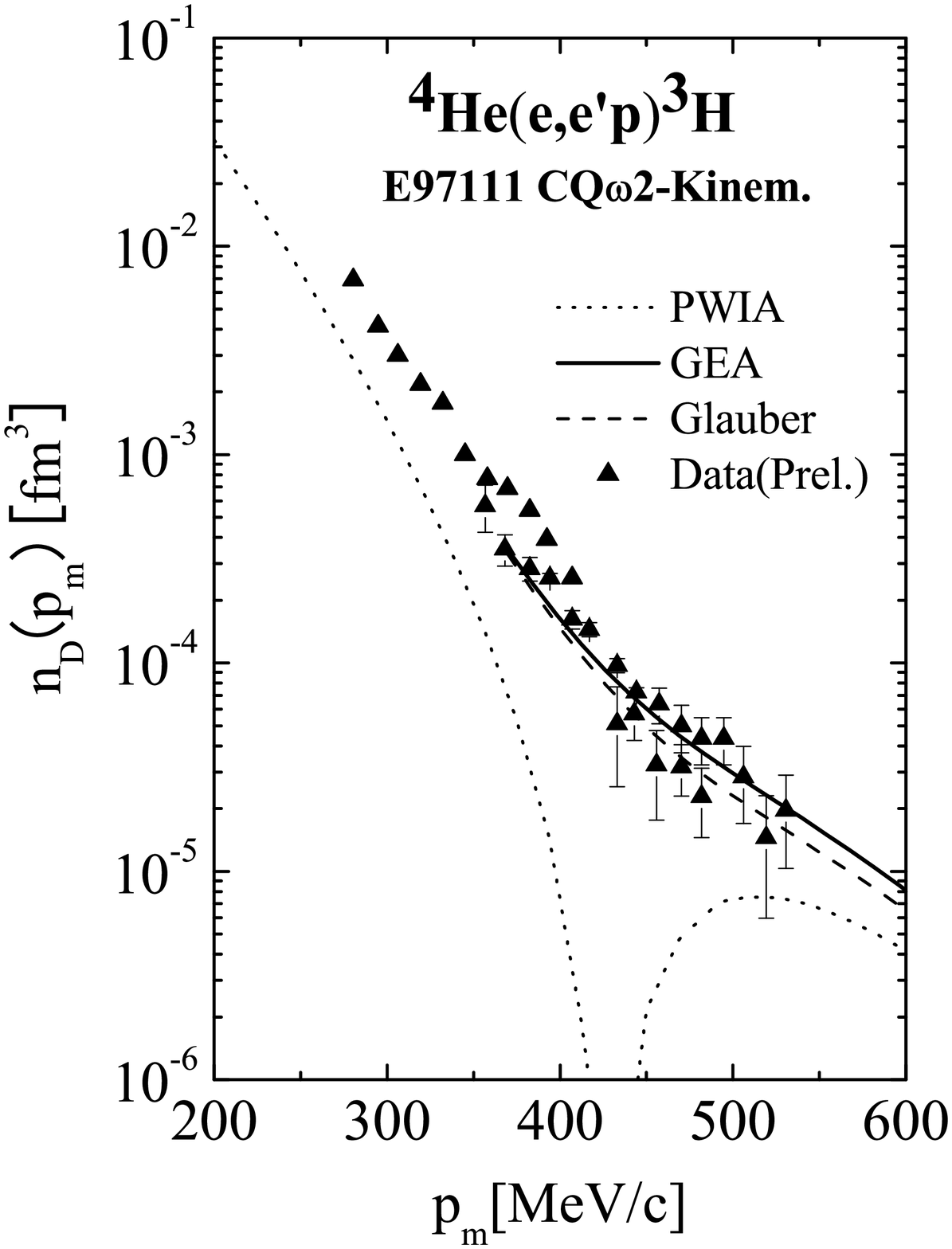}
    \epsfysize=8.0cm\epsfbox{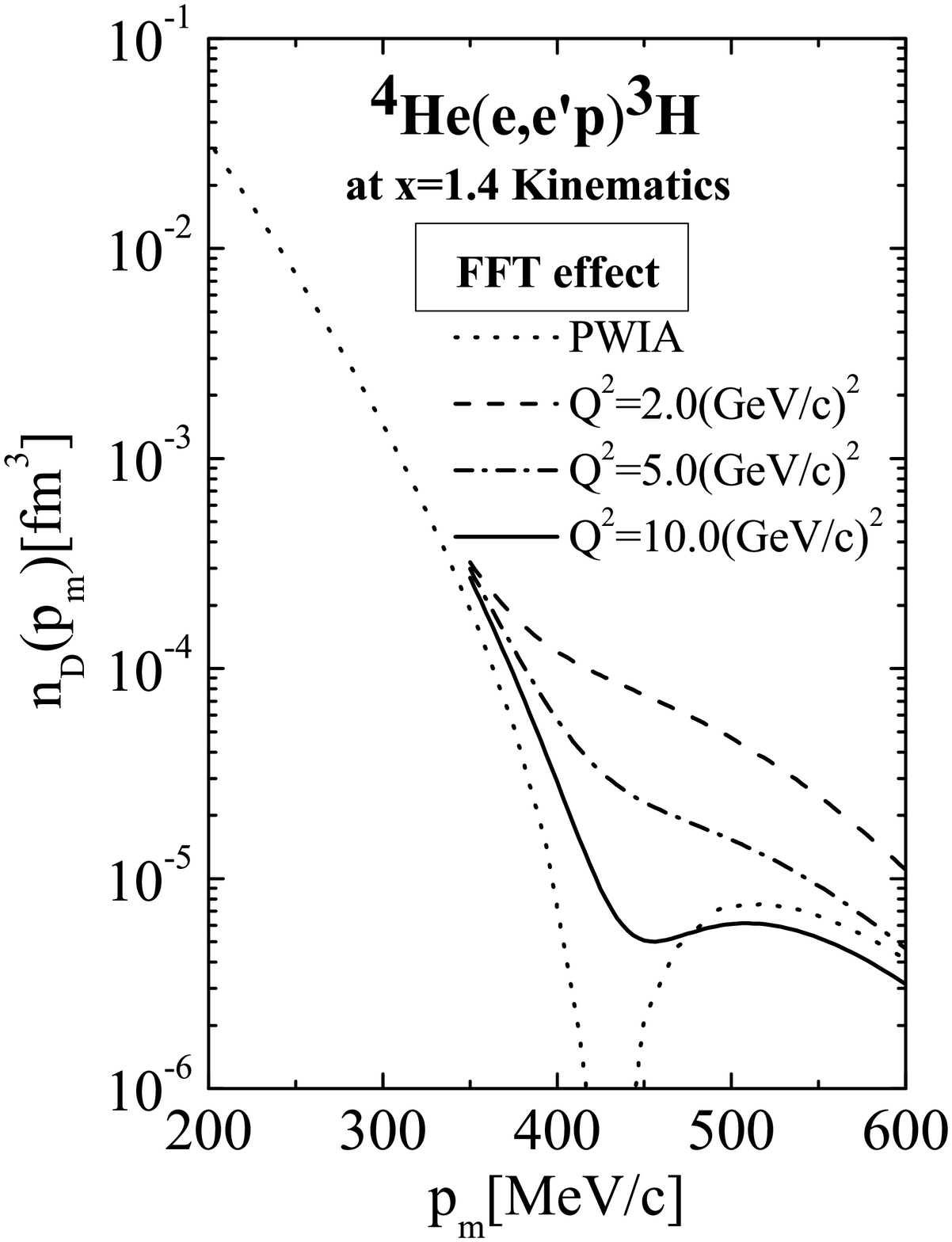}}
\caption{LEFT:\,\, the reduced cross section (Eq. (6))  of the
process $^4He(e,e'p)^3H$ at perpendicular
 kinematics and  $x \simeq 1.8$. The solid line shows the results within GEA, whereas
 the dashed curve corresponds to the conventional GA.
  Preliminary  data from \cite{E97111}.
RIGHT:\,\, the reduced cross section (Eq. (6))  of the process
$^4He(e,e'p)^3H$ at perpendicular
 kinematics for various values of $Q^2$ and  $x \simeq 1.4$, calculated taking FFT effects into
 account. Four-body wave functions from Ref. \cite{ATMS2}. (After Ref. \cite{CKM})}
  \label{Fig3}
\end{figure}

 The results for the 2bbu channel
 in  $^4He$,
    are reported
 in Figs. \ref{Fig3} where the reduced cross section
\begin{equation}
n_D(\mathbf{p}_m)=\frac{d^5 \sigma}
  {d \Omega ' d {E'} ~ d \Omega_p}
({\mathcal K} \sigma_{ep})^{-1},
\end{equation}
 is compared with preliminary  JLab data  (CQ$\omega$2) obtained in  perpendicular
  kinematics \cite{E97111}. It can be seen that: i) the dip predicted by the PWIA
 is completely  filled up  by the FSI; ii)  like the $^3He$ case, the
difference between GA and  GEA  is not very large; iii)   an
overall satisfactory agreement between theory and experiment is
obtained.

It has been argued by various authors that at high values of  $Q^2$ the
phenomenon of color transparency, i.e. a reduced NN cross section in the medium,
 might be observed. Color transparency is a consequence of the cancelation
 between various hadronic intermediate states of the produced ejectile.
 In \cite{Braun} the vanishing of FSI at $Q^2$ has been produced by considering the
 finite formation time (FFT) the ejectile needs to reach its asymptotic form of a physical baryon.
 This has been implemented by explicitly considering the  dependence
  of the NN scattering amplitude upon nucleon  virtuality.
According to \cite{Braun}  FFT effects can be introduced
by replacing
 $\theta(z_i-z_1)$ appearing in the Glauber profile with
 \be
 \textit{J}(z_i-z_1)= \theta(z_i-z_1)\left( 1-\exp [-(z_i-z_1)/{\it l}(Q^2)] \right)
 \label{eq:FFT}
 \ee
 \noindent where  ${\it l}(Q^2)={Q^2}/(x m_N\,M^2)$; here
  $x$ is the  Bjorken scaling  variable and the quantity $l(Q^2)$ plays
the role of the proton
 formation length, i.e. the length of the trajectory that the
knocked out proton runs until it return to its asymptotic form;
the quantity $M$ is related to the nucleon mass $m_N$ and to an
average  resonance state of mass $m^*$ by $M^2 = {m^*}^2 - m_N^2$.
 Since the formation length grows linearly with
$Q^2$, at higher $Q^2$ the strength of the Glauber-type FSI is
reduced by the damping factor $( 1-\exp[-(z_i-z_1)/{\it
l}(Q^2)])$;  if $l(Q^2)$ = $0$,
 then   $S_{FFT}$  reduces
to the usual Glauber operator $S_{G}$.
The results of calculations of the  cross section of the process
$^4He(e,e'p)^3H$ in perpendicular kinematics taking into account
FFT effects
 are presented in Fig.
\ref{Fig3} (for calculations in parallel kinematics see \cite{FFTHiko}).
It can be seen that at the JLAB
kinematics  ($Q^2=1.78\,(GeV/c)^2$, $x\sim1.8$) FFT effects, as expected,
 are too small to be detected, they can unambiguously be observed
in the region $5 \leq Q^2 \leq 10\, (GeV/c)^2$ and   $x$ =1.4.
 Thus measuring the  $Q^2$
dependence of the cross section of $^4He(e,e'p)^3H$ process at
$p_m\sim430\, MeV/c$ and  $Q^2\sim10\,(GeV/c)^2$ would be
extremely interesting.

To sum up, the following remarks are in order:

\noindent i) an overall good agreement between the results of
theoretical calculations
  and
 experimental data for both  $^3He$
 and $^4He$  is observed, which is very gratifying also
  in view of the lack of any adjustable parameter in theoretical calculations;
ii) the effects of the FSI are such that they  systematically
  bring theoretical calculations in better agreement with the experimental data;
 for some quantities, FSI effects  simply improve the agreement
 between theory and experiment, whereas for some other quantities,
 they play a dominant role;
iii)  the 3bbu channel in $^3He$,
  provides evidence of NN correlations, in that the experimental values of $p_m$ and $E_m$
   corresponding  to the
   maximum values of the cross section
satisfy,  to a large extent,  the relation  predicted by the
 two-nucleon correlation mechanism,    with   FSI mainly
 affecting  the magnitude of the cross section;
iv) the violation of the factorization approximation is
appreciable at "negative" values ($\phi =0$) of the missing momentum,
 whereas the non factorized and factorized predictions are
in good agreement in the whole range of positive values ($\phi
=\pi$) of $|\bp_m|$; v) the left-right asymmetry can
 reasonably be reproduced
at low values of the missing momentum, but a substantial discrepancy between
theoretical calculations and experimental data, common to several calculations,
 remains to be explained at high values of $|\bp_m|$;
vi)  Finite Formation Time   effects can be investigated at moderately high values of $Q^2$ by means
of the process  $^4He(e,e'p)^3H$.

\end{document}